\documentclass[a4paper,11pt]{article}   
\usepackage{fontenc,a4}

\usepackage{amsmath}
\usepackage{amsfonts}
\usepackage{amssymb}
\usepackage{graphicx}
\usepackage{pstricks}
\usepackage{multirow}
\usepackage{fancyhdr}
\usepackage[left=3cm,top=2.5cm,right=3cm]{geometry}
\usepackage{cite}

\newcounter{figures}
\newcounter{tables}
\newcommand{\athdm}[0]{A2HDM }
\newcommand{\athdmws}[0]{A2HDM}
\newcommand{\thdm}[0]{2HDM }
\newcommand{\thdmws}[0]{2HDM}
\newcommand{\e}{\mathrm{e}}
\newcommand{\cR}{\mathcal{R}}

\begin{document}
\begin{titlepage}
\begin{flushright} IFIC/10-34 \\  FTUV/10-1123 \end{flushright}
\vskip 2cm
\begin{center}
{\Large\bf The $\mathbf{\bar{B}\rightarrow X_s}\boldsymbol{\gamma}$ Rate and $\mathbf{CP}$ Asymmetry within \\[0.2cm] the Aligned Two-Higgs-Doublet Model}
\vskip 1cm
{\bf Martin Jung, Antonio Pich and Paula Tuz\'on}
\\[0.5cm]
Departament de F\'{\i}sica Te\`orica, IFIC, Universitat de Val\`encia - CSIC\\ Apt. Correus 22085, E-46071 Val\`encia, Spain
\\[0.5cm]
\today
\end{center}
\vskip 2cm

\begin{abstract}
In the two-Higgs-doublet model the alignment of the Yukawa matrices in flavour space guarantees the absence of flavour-changing neutral currents at tree level, while introducing new sources for $CP$ violation parametrized in a very economical way \cite{Pich:2009sp}. This implies potentially large influence in a number of processes, $b\to s\gamma$ being a prominent example where rather high experimental and theoretical precision meet. We analyze the $CP$ rate asymmetry in this inclusive decay and determine the resulting constraints on the model parameters. We demonstrate the compatibility with previously obtained limits \cite{Jung:2010ik}. Moreover we extend the phenomenological analysis of the branching ratio, and examine the influence of resulting correlations on the like-sign dimuon charge asymmetry in $B$ decays.
\end{abstract}
\thispagestyle{empty}
\vfill

\end{titlepage}

\section{Introduction}

Adding a second doublet to the scalar sector of the Standard Model (SM) has been
a popular idea since its proposal in the 70's \cite{Lee:1973iz},
due to its simplicity and being the low-energy limit of some more complete theories.
In the most general version of the model, the fermionic couplings of the neutral scalars are non-diagonal in flavour and, therefore, generate unwanted flavour-changing neutral-current (FCNC) phenomena.
Different ways to suppress these have been developed, giving rise to a variety of specific implementations \cite{Glashow:1976nt,Haber:1978jt,Hall:1981bc,Donoghue:1978cj,Barger:1989fj,Savage:1991qh,
Grossman:1994jb,Akeroyd:1998ui,Akeroyd:1996di,Akeroyd:1994ga,Aoki:2009ha,Ma:2008uza,Barbieri:2006dq,LopezHonorez:2006gr,Cheng:1987rs,Atwood:1996vj}. A new approach in that respect is given
by the Aligned Two-Higgs-Doublet Model (A2HDM) \cite{Pich:2009sp},
where each pair of Yukawa coupling matrices
(describing the couplings to right-handed up-type quarks, down-type quarks and leptons, respectively)
is assumed to be aligned in flavour space.
The resulting Yukawa Lagrangian \cite{Pich:2009sp},
\begin{eqnarray}\label{lagrangian}
\mathcal L_Y &\! =&\! - \frac{\sqrt{2}}{v}\, H^+(x) \left\{  \bar{u}(x)   \left[  \varsigma_d\, V M_d \mathcal P_R - \varsigma_u\, M_u^\dagger V \mathcal P_L  \right]   d(x)\, +\, \varsigma_l\; \bar{\nu}(x) M_l \mathcal P_R l(x)\right\} \nonumber \\ & &\! -\,\frac{1}{v}\; \sum_{\varphi, f}\, y^{\varphi^0_i}_f\, \varphi^0_i(x)  \; \bar{f}(x)\,  M_f \mathcal P_R  f(x)\;  + \;\mathrm{h.c.} \; ,
\end{eqnarray}
contains three complex couplings $\varsigma_f$ ($f=u,d,l$) which parametrize all possible freedom allowed by the alignment conditions.
$\mathcal P_{R,L}\equiv \frac{1\pm \gamma_5}{2}$ are the right-handed and left-handed projectors and the  couplings of the three neutral scalar fields
$\varphi^0_i(x) =\{ h(x), H(x), A(x)\}$ 
are given in terms of $\varsigma_f$ and the orthogonal transformation
$\cR$ relating the scalar weak and mass eigenstates:
$y_{d,l}^{\varphi^0_i} = \cR_{i1} + (\cR_{i2} + i\,\cR_{i3})\,\varsigma_{d,l}$, \
$y_u^{\varphi^0_i} = \cR_{i1} + (\cR_{i2} -i\,\cR_{i3}) \,\varsigma_{u}^*$.

While the CKM matrix $V$ remains the only source of flavour-changing phenomena, the
universal (flavour-blind) couplings $\varsigma_f$ introduce three new complex phases
and, therefore, a new source of $CP$ violation. For particular (real) values of these parameters the usual $CP$-conserving models based on discrete $\mathcal{Z}_2$ symmetries
are recovered.
Quantum corrections induce a misalignment of the Yukawa matrices, generating small
FCNC effects \cite{Pich:2009sp,Jung:2010ik,Ferreira:2010xe,Braeuninger:2010td}. The flavour symmetries of the A2HDM strongly constrain the allowed FCNC
structures, providing at the quantum level \cite{Pich:2009sp,Jung:2010ik}
an explicit implementation of the popular Minimal Flavour Violation (MFV) scenarios
\cite{Chivukula:1987py,Hall:1990ac,Buras:2000dm,D'Ambrosio:2002ex,Cirigliano:2005ck,Kagan:2009bn,Buras:2010mh,Trott:2010iz},
but allowing at the same time for new $CP$-violating phases.

A MFV structure within the context of the type II two-Higgs doublet model (\thdmws) was considered in \cite{D'Ambrosio:2002ex} and flavour-blind phases 
have been recently incorporated in \cite{Buras:2010mh}\footnote{Regarding MFV and $CP$ violating phases, see also \cite{Colangelo:2008qp,Kagan:2009bn,Botella:2009pq,Paradisi:2009ey,Feldmann:2009dc}.}. 
These references perform a perturbative expansion around the usual $U(1)_{PQ}$ symmetry limit (type II)
and look for $\tan{\beta}$--enhanced effects.
Since the A2HDM does not assume any starting ad-hoc symmetry, it leads to a
more general MFV framework with $\tan{\beta}$ substituted
by the 
six-dimensional
parameter space spanned by the couplings $\varsigma_f$.
While giving rise to a much richer phenomenology, the A2HDM implies
an interesting hierarchy of FCNC effects, avoiding the stringent experimental constraints for light-quark systems and allowing at the same time for potentially relevant signals in
heavy-quark transitions \cite{Jung:2010ik}. Notice that in the general case, without
$U(1)_{PQ}$ or $\mathcal{Z}_2$ symmetries, $\tan{\beta}$ does not have any physical
meaning because it can be changed at will through $SU(2)$ field redefinitions in the scalar space; the physics needs to be described through the (scalar-basis independent)
parameters $\varsigma_f$.

The presence of new weak phases immediately poses the question of compatibility with existing measurements of $CP$ violating observables. One decay where a high sensitivity
is expected is $b\to s\gamma$, for three reasons: the SM asymmetry is known to be tiny \cite{Ali:1998rr,Hurth:2003dk,Kagan:1998bh}, its measurement is rather precise and compatible with zero, and the potential influence of new physics in this decay is large. In addition, being an inclusive process, non-perturbative uncertainties are expected to be under better control than for exclusive modes.
The branching ratio and the $CP$ rate asymmetry have both been analyzed within the context of a general \thdm \cite{Borzumati:1998tg}, and it has been pointed out that in large regions
of the parameter space the NLO predictions are not reliable, because they suffer from
a very large renormalization-scale dependence.
For the part of the parameter space not affected by these problems the asymmetry was found to be small.

In this work, we update the calculation for the asymmetry within the framework of the \athdmws, extending the phenomenological analysis and discussing its implications for the model parameters. We will show that the combined phenomenological constraints from  $Z\to\bar{b}b$, $B^0$-mixing, $K^0$-mixing and Br($b\to s\gamma$) \cite{Jung:2010ik} exclude the
problematic regions of the parameter space. The predicted 
range for the 
$CP$ asymmetry turns out to be within the present experimental limits for all allowed ranges of the \athdm couplings, although for
some particular values of the parameters it is not far from the achieved sensitivity.
This makes future high-precision measurements of this quantity very desirable and calls for
a better control of theoretical uncertainties.

To that end, we first analyze the branching ratio in some more detail. 
We have started to discuss this observable in \cite{Jung:2010ik} together with others, but deem a more detailed analysis in this context worthwhile, due to its relevance for the rate asymmetry, but also due to its own high sensitivity and its intimate relation to the like-sign dimuon charge asymmetry (LDCA) in $B^0$-decays.

We proceed as follows: In section~\ref{bs} 
we discuss 
the $\bar{B}\rightarrow X_s \gamma$ branching-ratio prediction within the \athdmws,  
extending our results
derived in \cite{Jung:2010ik}. In section~\ref{rate} we analyze the $CP$ rate asymmetry, with emphasis on its scale dependence, and discuss its potential impact on the parameter space of the \athdmws. 
Finally, we conclude in section~\ref{conclusions} with a brief summary of our findings, 
and analyze their impact regarding the possible influence of charged-scalar effects on the LDCA in our framework.

\section{$\bar{B}\rightarrow X_s \gamma$ branching ratio}\label{bs}

Because of its importance for the following discussion, we first 
perform an
analysis of the $\bar{B}\rightarrow X_s \gamma$ branching ratio.
In the SM, this decay is known to be subject to large radiative corrections. While the problem of a sizable scale dependence has been basically resolved with the calculation of the NLO corrections, the issue of charm-mass scheme dependence can only be addressed at NNLO \cite{Greub:1996jd}. To achieve this, a huge effort is being made,
and by now the branching ratio is
essentially\footnote{For the present status and recent developments, see e.g. \cite{Misiak:2010dz,Paz:2010wu,Hurth:2010tk}.}
calculated up to NNLO in the 
SM. We follow here the calculation by \cite{Misiak:2006zs}, giving $\mathrm{Br}(\bar{B}\rightarrow X_s \gamma)^{\rm SM,theo}_{E_{\gamma}>1.6\, {\rm GeV}}=(3.15\pm 0.23)\times 10^{-4}$,
in agreement with the present world average \cite{HFAG:2010qj}
\begin{equation}
\mathrm{Br}(\bar{B}\to X_s\gamma)_{E_\gamma\ge1.6\, {\rm GeV}}^{\rm exp}=(3.55\pm0.26)\times10^{-4}\,.
\end{equation}
Note however, that different treatments of photon-energy-cut related effects lead to slightly different results\cite{Becher:2006pu,Andersen:2006hr,Ligeti:2008ac} (see also \cite{Misiak:2008ss}). The related shifts are of the order of the uncertainty assigned in \cite{Misiak:2006zs}. Regarding the non-perturbative part of this calculation, contributions with the photon coupling to light partons (``resolved'' photon contributions) lead to the appearance of non-local matrix elements, implying an unreducible error of $\sim5\%$ \cite{Benzke:2010js}.

In the \thdmws, the calculation has been performed up to NLO
\cite{Degrassi:2010ne, Ciuchini:1997xe, Borzumati:1998tg, Ciafaloni:1997un}.
For $m_s=0$, the effective low-energy operator basis remains the same as in the SM and the modifications induced by new-physics contributions
appear only in the Wilson coefficients:
\begin{equation}\label{eq::Cstructure}
C_i^{\mathrm{eff}}(\mu_W)=C_{i,SM}+|\varsigma_u|^2\; C_{i,uu}-(\varsigma_u^*\varsigma_d)\; C_{i,ud} \; ,
\end{equation}
where $\varsigma_u^*\varsigma_d{\phantom{*}}=|\varsigma_u||\varsigma_d|\, \e^{i\varphi}$ defines the relative phase $\varphi$. Values around $\varphi=0\; (\pi)$ correspond to destructive (constructive) interference with the SM amplitude, as occurs in the \thdm of type I (II).

Since the 2HDM contribution to the decay amplitude has only been calculated up to NLO, the corresponding terms of $\mathcal O(\alpha_s^2)$ are neglected consistently in the rate.
An important observation in \cite{Borzumati:1998tg} is that the stabilization against scale variations does not work as well for the general \thdm as for the SM,
because in some regions of the parameter space
cancellations between SM and new-physics contributions occur, enhancing the sensitivity to
higher-order QCD corrections.
In \cite{Jung:2010ik} we have used the NNLO SM calculation,
retaining terms up to order $\alpha_{s,SM}^2$ and $\alpha_{s,NP}$
which stabilize the result, however with some problematic regions, leading to unphysical
results, still remaining. But these regions (corresponding to large values of $|\varsigma_u|$) are excluded by constraints coming from $Z \rightarrow \bar{b}b$, $\epsilon_K$ and $\Delta m_{B_s^0}$, allowing us to perform a consistent theoretical study of
$\mathrm{Br}(\bar{B}\to X_s\gamma)$ in the whole remaining parameter space.

\begin{figure}[tbh!]
\begin{center}
\includegraphics[width=9cm]{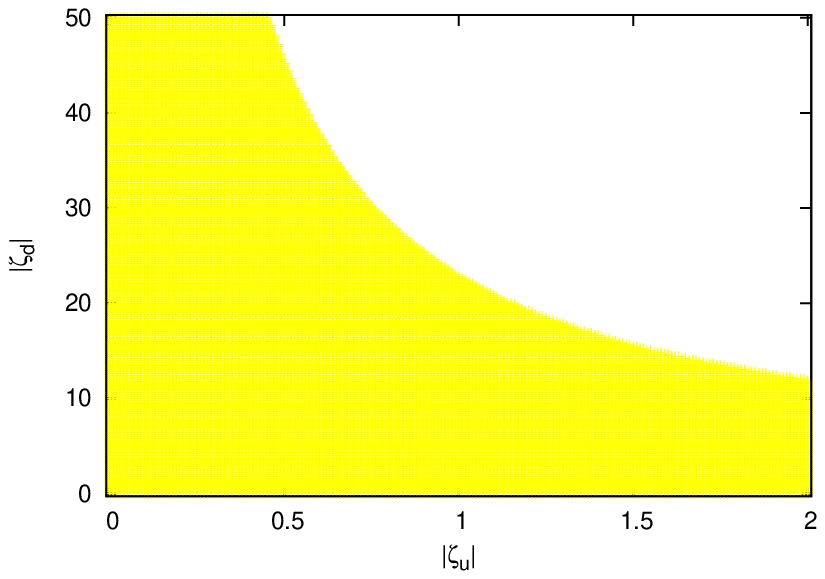}
\caption{\it\label{fig::udold}
The constraint from ${\rm Br}(b\to s\gamma)$ plotted in the plane $|\varsigma_u|-|\varsigma_d|$, see text.}
\end{center}
\end{figure}

In Fig.~\ref{fig::udold} we show the resulting constraint in the plane $|\varsigma_u|-|\varsigma_d|$, varying the remaining parameters in the ranges $M_{H^\pm}\in[80,500]$~GeV and $\varphi\in[0,2\pi]$, and treating the errors as described in \cite{Jung:2010ik}. Due to the additional degrees of freedom, the resulting constraint is relatively weak; especially, no direct bound can be found with respect to the charged-scalar mass, in striking contrast to the \thdm type II. As discussed in \cite{Jung:2010ik}, this changes as soon as the phase is kept fixed to a certain value.

\begin{figure}[tbh!]
\begin{center}
\includegraphics[width=7cm]{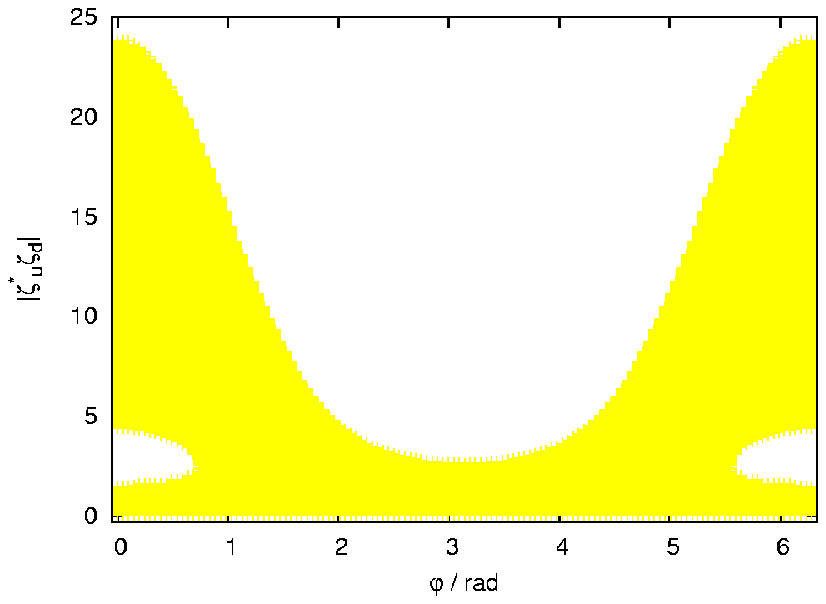} \qquad \includegraphics[width=7cm]{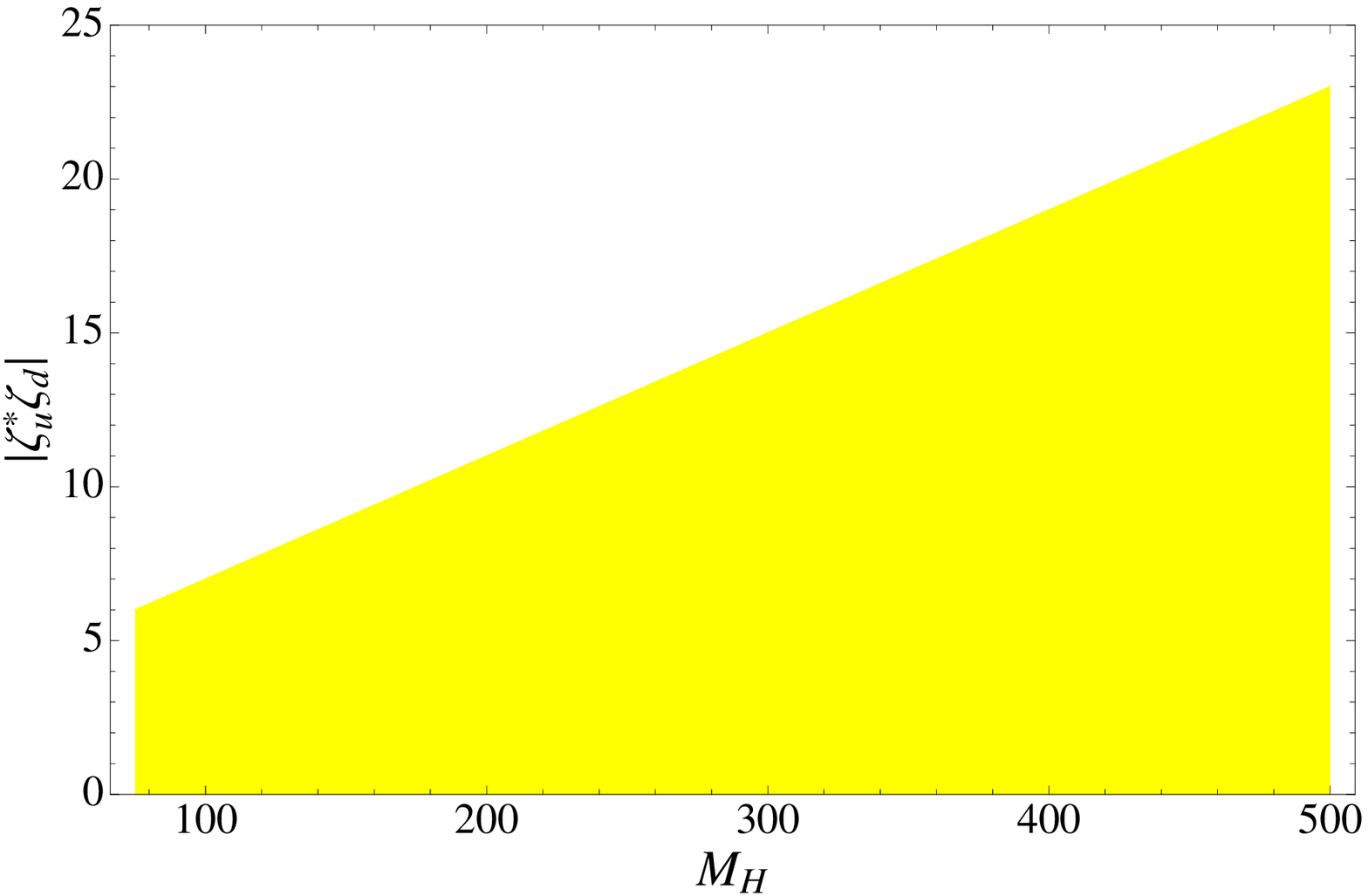}
\caption{\it\label{fig::BRcorrelations}
The constraint from ${\rm Br}(b\to s\gamma)$ plotted in the planes $|\varsigma_u^*\varsigma_d|-\varphi$ (left) and $|\varsigma_u^*\varsigma_d|-M_{H^\pm}$ (right).}
\end{center}
\end{figure}

Numerically, the decay amplitude has roughly the following structure for large scalar masses (using numerical estimates for the different contributions from \cite{Borzumati:1998tg}):
\begin{equation}\label{eq::numerics}
A\sim A_{SM} \left\{  1 - 0.1\; \varsigma_u^* \varsigma_d\left(\frac{500~\rm{GeV}}{M_{H^\pm}}\right)^2 + 0.01\; |\varsigma_u|^2\left(\frac{500~\rm{GeV}}{M_{H^\pm}}\right)^2\right\} \; .
\end{equation}
From this it is obvious that we can expect in that case constraints on the parameter combinations $|\varsigma_u|^2/M_{H^{\pm}}^2$ and $\varsigma_u^*\varsigma_d/M_{H^{\pm}}^2$, the latter being complex.
For sizable $|\varsigma_d|$, the last term is negligible, as $|\varsigma_u|$ is constrained to be $\mathcal{O}(1)$ at most, leaving only the dependence on the combination $\varsigma_u^*\varsigma_d/M_{H^{\pm}}^2$. The resulting limits on the single parameters in this ratio are relatively weak, as seen above. The strength of the constraint lies however in creating strong correlations between different parameter combinations. This is illustrated in Fig.~\ref{fig::BRcorrelations}, where the  product $|\varsigma_u^*\varsigma_d|$ is plotted against the charged-scalar mass $M_{H^\pm}$ and the relative phase $\varphi$, respectively. 
The hole in the left plot can be understood as separating two regions where the NP influence is relatively small 
(lower part) and 
where it is approximately twice the size of the SM contribution 
(upper part), the latter corresponding to a fine-tuned solution. For larger phases it is not possible anymore to cancel the SM amplitude completely, so only one constraint is remaining, making the gap disappear. From the right plot it becomes obvious that high values for the product $|\varsigma_u^*\varsigma_d|$ are only allowed for large values of the charged-scalar mass, due to the restriction on the ratio. 
The effect of these correlations can obviously be large; it is taken into account in the following analysis.

\section{$\mathbf{CP}$ rate asymmetry}\label{rate}

The $CP$ rate asymmetry for the process $b\to s\gamma$ is defined as
\begin{equation}\label{acpformula}
a_{CP}= \frac{\mathrm{Br}(\bar{B}\to X_{s}\gamma)-\mathrm{Br}(B\to X_{\bar{s}}\gamma)}{\mathrm{Br}(\bar{B}\to X_s\gamma)+\mathrm{Br}(B\to X_{\bar{s}}\gamma)} \,.
\end{equation}
Being doubly Cabibbo suppressed, it is known to be tiny in the SM, below $1\%$ \cite{Ali:1998rr,Kagan:1998bh,Hurth:2003dk}, making it a sensitive probe of new-physics effects. The relative scale uncertainty is large, as it is to be expected from a NLO calculation for a $CP$ asymmetry, because it is the first contributing order. However, it is small in absolute terms. The same is true for the charm-mass dependence. Note that the enhanced power-corrections at order $\mathcal{O}(\alpha_s\frac{\Lambda}{m_b})$ \cite{Benzke:2010js}, mentioned above in the discussion of the rate, are not yet included in the calculation for the asymmetry. Being of relative order $\Lambda/m_b$, they are likely to increase the uncertainty significantly. 
However, the experimental uncertainty of the present world average \cite{HFAG:2010qj},
\begin{equation}
a_{CP}^{\mathrm{exp}}(\bar{B}\rightarrow X_s \gamma)=-0.012\pm 0.028\,,
\end{equation}
is much larger than the expected value, so the calculation at NNLO has been considered less interesting up to now.

Within the context of the general \thdm the analysis has been performed at NLO
\cite{Borzumati:1998tg}, working in the limit $V_{ub}^{\phantom{'}}V_{us}^*=0$, which implies by unitarity that $V_{cb}^{\phantom{.}}V_{cs}^*=-V_{tb}^{\phantom{'}}V_{ts}^*$.
It was pointed out in \cite{Borzumati:1998tg} that, given
the cancellation problems in the branching ratio described before,
the prediction was not reliable in part of the parameter space where it exhibited a large scale dependence, but that
for parameter choices where the branching ratio was well behaved the predicted $a_{CP}$ was very small, $\mathcal{O}(1\%)$.

We now turn to reanalyze the asymmetry in the context of the \athdmws. In our previous study we have used the SM NNLO calculation of the branching ratio; however, we will not include NNLO
information in the predicted asymmetry, for two reasons: First,
the necessary calculation does not exist completely,
and the interpolation for the charm-mass dependence has been done specifically for the branching ratio, ignoring some imaginary parts which can be relevant for the asymmetry.
Second, and more importantly, in our approximation the asymmetry is a pure new-physics effect. Therefore, QCD corrections should enter at the same level for the SM and new-physics amplitudes. Moreover, as the scale dependence at NLO is in general more severe for the \athdm
than for the SM, inclusion of the NNLO SM corrections only would not stabilize the full result.

We confirmed these expectations explicitly by calculating the asymmetry with the NNLO SM contributions to the branching ratio included. The result showed a relatively large shift in the central value, which we attribute to the different charm-mass dependence, and no stabilization of the scale dependence at all. This behaviour is uniform
for different values of $\varsigma_{u,d}$ and $M_{H^\pm}$. We conclude that for the inclusion of NNLO corrections important parts are missing, the calculation of which
is beyond the scope of this paper.
This lack of NNLO corrections will make the problems described in the branching-ratio analysis reappear. Nevertheless, we start by analyzing the asymmetry for the same range of parameters considered in our previous study of the branching ratio, $|\varsigma_u|\in[0,2]$, $|\varsigma_d|\in[0,50]$, $\varphi\in[0,2\pi]$, and $M_{H^\pm}\in[80,500]$~GeV, to examine the strength of this observable by itself.

\begin{figure}[tbh!]
\begin{center}
\includegraphics[width=7cm]{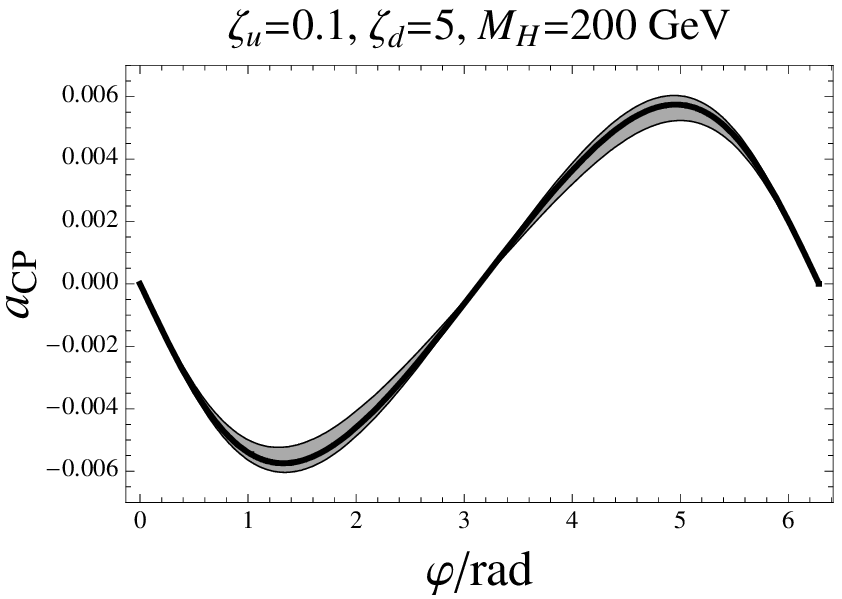} \qquad \includegraphics[width=7cm]{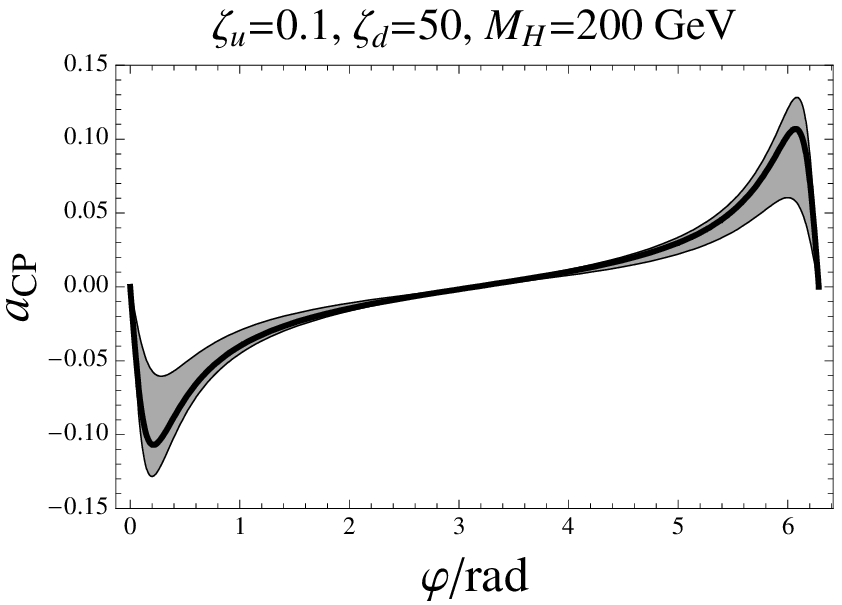}
\caption{\it
$CP$ asymmetry at NLO, as function of the relative phase $\varphi$, for $\varsigma_u=0.1$, $\varsigma_d=5$ (left) and $50$ (right) and $M_{H^\pm}=200$~GeV.
The band shows the variation with the scale, taking
$\mu_b=2.5$~GeV (central black line), 2~GeV (outer line) and 5~GeV (inner line).\label{mub}}
\end{center}
\end{figure}

Figure~\ref{mub} shows the $\mu_b$ renormalization-scale dependence of the predicted $CP$ asymmetry at NLO, as a function of the phase $\varphi$, taking $M_{H^\pm}=200$~GeV,
$|\varsigma_u|=0.1$ and $|\varsigma_d|=5$ (left) and $50$ (right).
The black central curve represents $a_{CP}$ for $\mu_b=2.5$~GeV, while
the outer (larger absolute values) and inner (smaller absolute values) lines correspond to $\mu_b=2$ and 5~GeV, respectively, 
using the same range of variation considered in the branching ratio analysis.
We observe that the $\mu_b$ dependence is proportional to $|a_{CP}|$; therefore in the regions where the asymmetry is relevant the theoretical error from the scale dependence is as well. As noted before, the overall scale dependence is strong, approximately $25\%$ in most of the parameter space. If not noted otherwise, in the following the scale is fixed to $\mu_b=2.5$~GeV.

\begin{figure}[tbh!]
\begin{center}
\includegraphics[width=9cm]{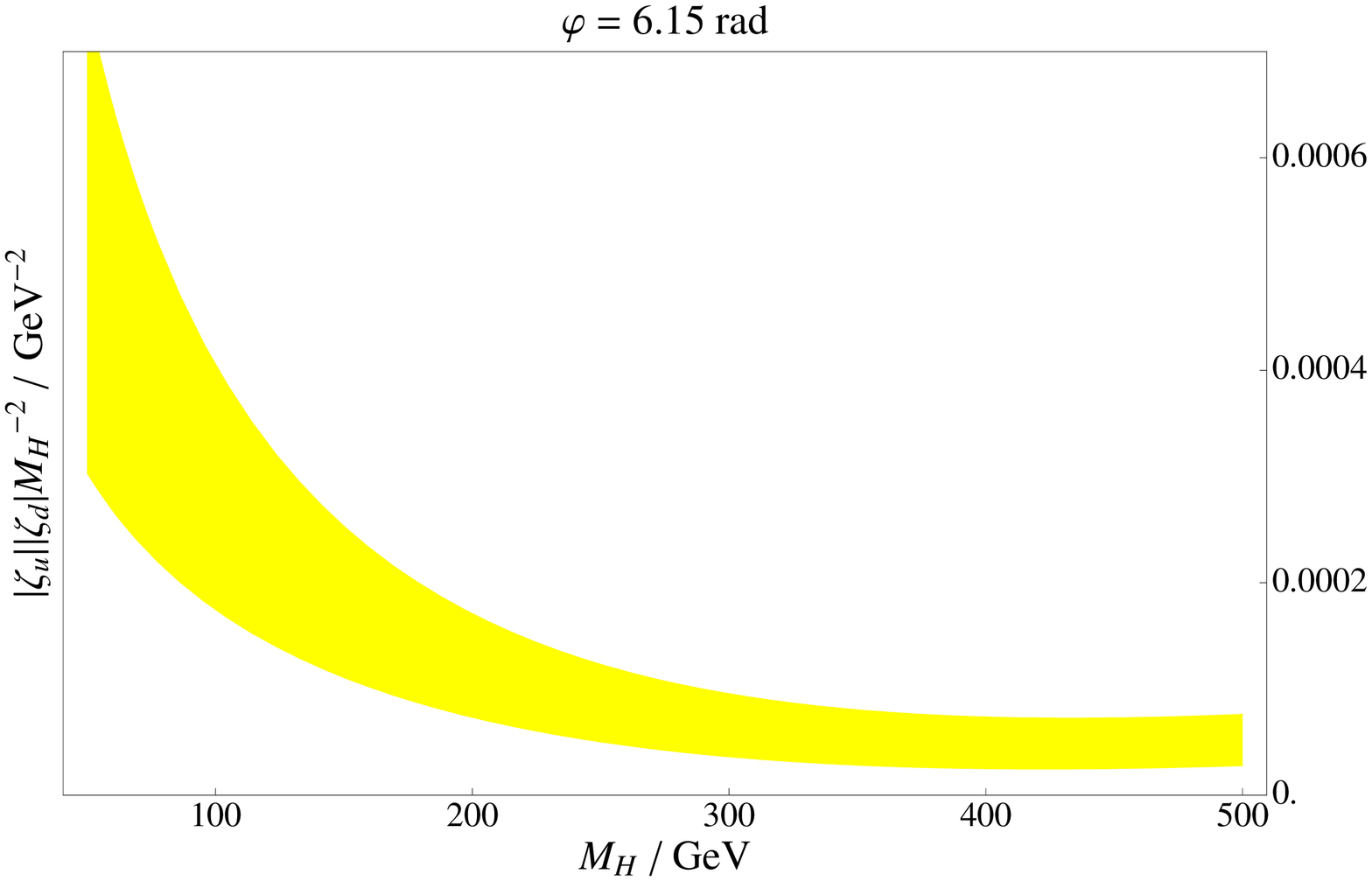}
\caption{\it
Region in the $|\varsigma_u||\varsigma_d|/M_{H^\pm}^2-M_{H^\pm}$ plane leading to values of $a_{CP}$ outside the (95\% CL) allowed experimental range, for fixed phase $\varphi=6.15$~rad and scale $\mu_b=2.5$~GeV ($M_{H^\pm}$ in GeV units).
\label{res}}
\end{center}
\end{figure}

The asymmetry is small for most values of the phase $\varphi$, apart from values around zero,
where for some parameter choices one observes an unphysical enhancement of $|a_{CP}|$.
This is apparent in Fig.~\ref{mub} (right panel), corresponding to $\varsigma_d = 50$,
which shows a large predicted asymmetry around $\varphi = \pm 0.2$~rad (modulo $2\pi$), with a very
large scale dependence.
This enhancement does not correspond to any large $CP$ contribution, but rather to a
destructive interference in the decay amplitude leading to a nearly vanishing rate,
which is only possible for small imaginary parts. This resonant behaviour can 
again be easily
understood from Eq.~\ref{eq::numerics}; a zero of the amplitude occurs around $\varsigma_u^* \varsigma_d/M_{H^\pm}^2\sim 4\times 10^{-5}\; {\rm GeV}^{-2}$ with $\e^{i\varphi}\sim 1$.
For lower scalar masses the $|\varsigma_u|^2$ term can give a more sizable contribution,
modifying this simplified behaviour, but again 
the amplitude can become small
through the cancellation of the SM and new-physics contributions. We illustrate this fact in
Fig.~\ref{res}, which shows the parameter region in the
$|\varsigma_u||\varsigma_d|/M_{H^\pm}^2-M_{H^\pm}$ plane leading to values of $a_{CP}$ outside the (95\% CL) allowed experimental range, taking  $\varphi=6.15$~rad and fixing the renormalization scale at $\mu_b=2.5$~GeV.

The regions of small amplitudes
around values of $\varphi \sim 0, 2\pi$ are of course
very sensitive to the adopted truncation of the perturbative expansion in powers of the strong coupling and introduce a correspondingly large theoretical uncertainty. Their appearance signals the need to incorporate higher-order corrections into the calculation. Fortunately, since the measured branching ratio is definitely non zero (by many standard deviations)
and compatible with the SM prediction, these
problematic regions of the parameter space, where cancellations occur, are already excluded.
Imposing the constraint that the experimental branching ratio should be reproduced, the
surviving allowed parameter ranges lead to well-behaved amplitudes and small asymmetries below the present limits.

\begin{figure}[tbh!]
\begin{center}
\includegraphics[width=9.cm]{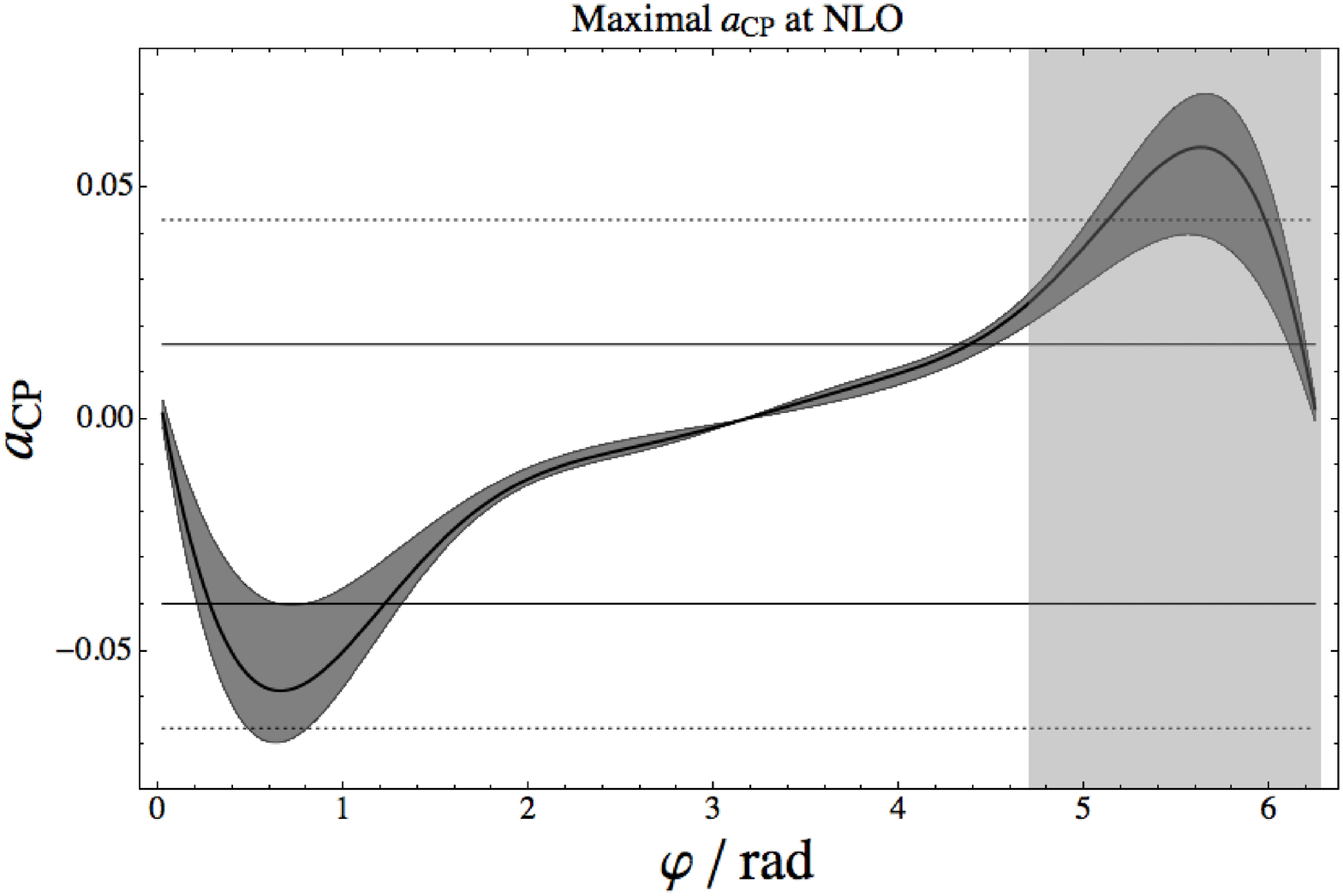}
\caption{\it
Maximal value of $a_{CP}$ versus the phase $\varphi$ at NLO, for
$M_{H^\pm}\in [80,500]$~GeV, $|\varsigma_u| \in [0,2]$ and $|\varsigma_d| \in [0,50]$,
taking into account the experimental Br($\bar{B} \rightarrow X_s \gamma$) constraint at $95\%$ CL. The three curves correspond to
$\mu_b=2$ (outer), $2.5$ (center) and $5$~GeV (inner).
The dotted (continuous) horizontal lines denote the allowed experimental range at
$95\%$~CL (68\% CL).
The vertical shadowed band indicates the region in which an enhancement of the LDCA induced by charged-scalar effects in the direction of the measured value may occur.
\label{acpbsgamma}}
\end{center}
\end{figure}

In Fig.~\ref{acpbsgamma} we plot the predicted maximal value for the $CP$ rate asymmetry,
as a function of the Yukawa phase $\varphi =\arg (\varsigma_u^*\varsigma_d^{\phantom{*}})$,
including the Br($\bar B \to X_s \gamma$) constraint by showing only points that lie within its $95\%$~CL experimental range. We have scanned the predictions over the parameter ranges
$M_{H^\pm}\in [80,500]$~GeV, $|\varsigma_u| \in [0,2]$ and $|\varsigma_d| \in [0,50]$, and have treated the error for the branching ratio as explained in \cite{Jung:2010ik}.
We find as expected that the problematic regions have been excluded, and that the maximal achievable asymmetry is compatible with the present experimental limits at $95\%$~CL,
given the scale dependence of the prediction.
Sizable asymmetries at the 1-5\% level seem possible for $\varphi\sim\pm 0.7$~rad,
but the corresponding theoretical uncertainty is unfortunately quite large, as shown
by the large scale-dependence of the theoretical results.

\section{Discussion}\label{conclusions}

The \athdm introduces new sources of $CP$ violation in the flavour sector while avoiding FCNCs at tree level, and provides an explicit counter-example to the widespread assumption that
in \thdmws s without tree-level-FCNCs all $CP$-violating phenomena should originate in the CKM matrix. 
Since all Yukawa couplings are proportional to fermion masses, the \athdm gives rise to an interesting hierarchy of FCNC effects, avoiding the stringent experimental constraints for light-quark systems and allowing 
at the same time 
for
interesting signals in heavy-quark transitions. The flavour-blind phases present in the model imply a broad range
of very interesting phenomenological implications, opening new possibilities
which are worth to be investigated.

An obvious question to address is the phenomenologically allowed size for the new phases,
which should be constrained by the measured $CP$-violating observables in flavour physics. Of these, the $b\to s\gamma$ asymmetry is known to have both a very small SM contribution and a high sensitivity to new physics. Unfortunately, although the $b\to s\gamma$ 
rate
is known at NNLO in the SM, the \thdm contributions have been only computed at NLO. Since a non-zero rate asymmetry requires absorptive contributions leading to a strong-phase difference, it first appears at NLO. Therefore, the theoretical uncertainties are much larger than for the branching ratio. On the other hand, some of the parametric uncertainties cancel in the asymmetry.

We have analyzed the predicted $b\to s\gamma$ rate asymmetry within the \athdmws, scanning the
parameter space of the model over the full domain allowed by known constraints
from other processes, which we have derived in a previous work \cite{Jung:2010ik}.
We have shown that, while $a_{CP}$ could be enhanced for some particular values of the parameters, it remains compatible with its present measurement when considering parameters leading to acceptable values for the branching ratio.
Sizable $CP$ asymmetries at the 1-5\% level, close to the present experimental bound, seem possible for Yukawa phases around
$\arg (\varsigma_u^*\varsigma_d)\sim\pm 0.7$~rad,
making a future measurement of this quantity very interesting. Such a measurement would be possible at a Super-$B$ factory, where precisions better than $1\%$ could be achieved \cite{Bona:2007qt,Akeroyd:2004mj}.
However, the presently large theoretical uncertainty from scale and mass dependence makes necessary a complete NNLO calculation for the SM and the \athdm contributions to fully exploit such a measurement. At the 1\% level, one also needs to analyze the small corrections induced by $|V_{ub}V_{us}^*|\neq 0$, which have been neglected up to now. An exception is given by potential experimental information on the quadrant of the new-physics weak phase $\varphi$: once a measurement is available which determines the sign of the asymmetry unambiguously, two quadrants are excluded.

We also extended the phenomenological analysis of the branching ratio over the first assessment done
in our previous publication \cite{Jung:2010ik}. 
The strong constraints on the model parameters e.g. in type II models are not visible in the \athdmws, due to the presence of additional parameters. They are replaced by correspondingly strong correlations, which we analyzed in some detail above, and which play an important role when combining other constraints with this observable.

\begin{figure}[tbh!]
\begin{center}
\includegraphics[width=7.cm]{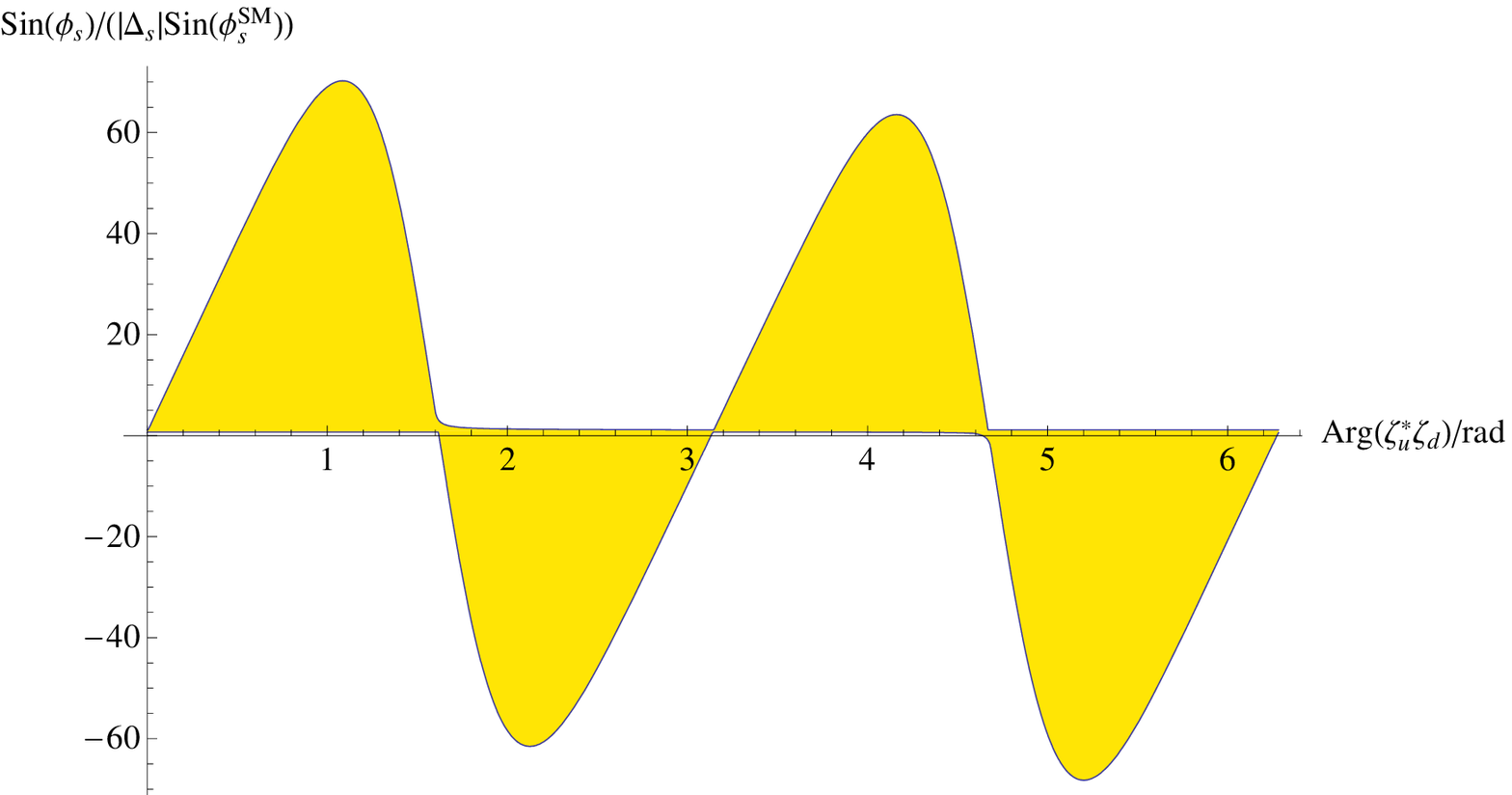}\qquad\includegraphics[width=7.cm]{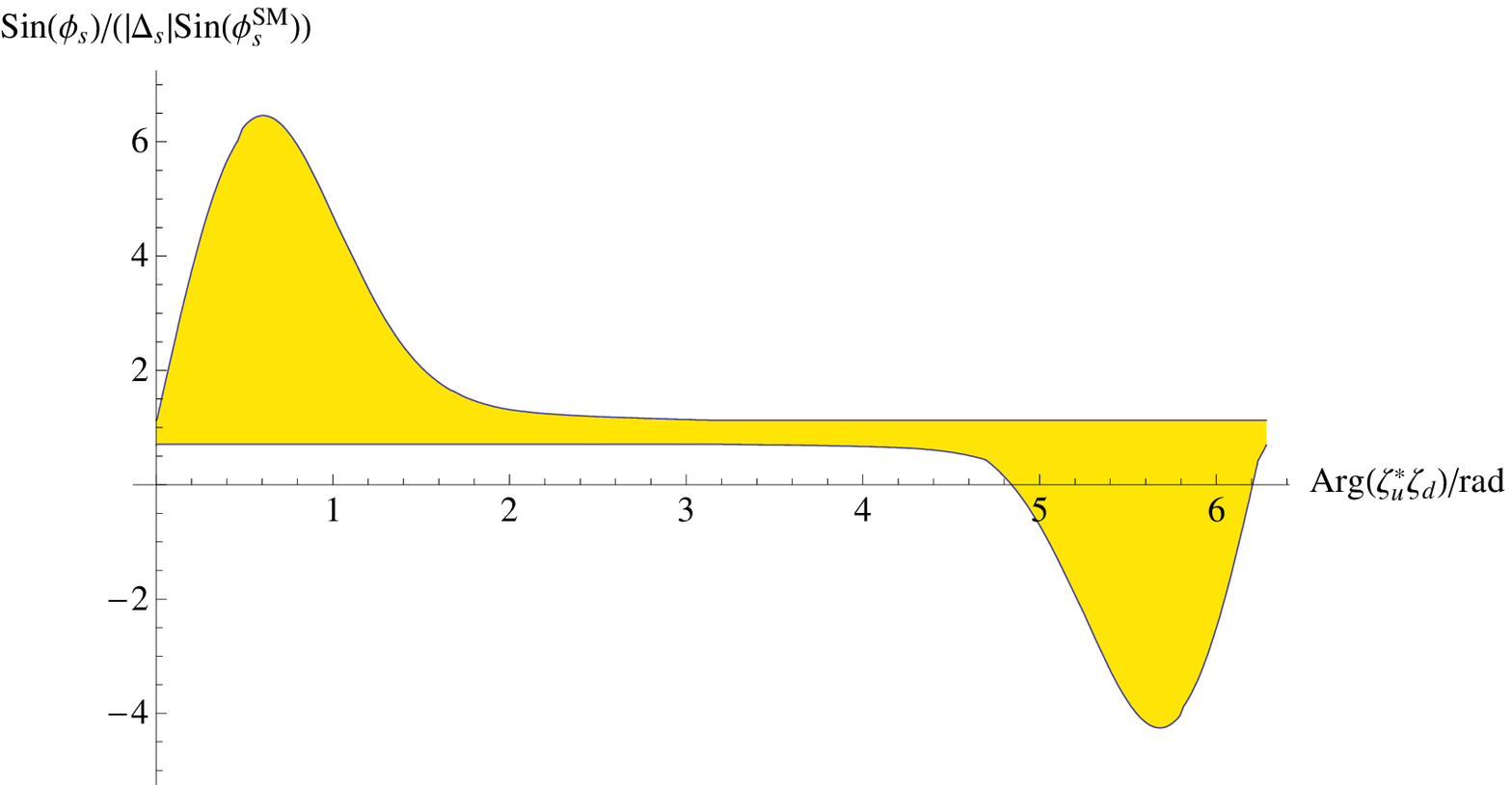}
\caption{\it\label{fig::LDCAnew}
Possible influence of charged-scalar-effects on the LDCA, taking the constraint from ${\rm Br}(b\to s\gamma)$ shown in Fig.~\ref{fig::udold} (left) and Figs~\ref{fig::BRcorrelations} (right) into account.  Shown is the allowed range for the relative factor with respect to the SM result as a function of the phase $\varphi$; see \cite{Jung:2010ik} for details of the calculation.
}
\end{center}
\end{figure}

An example for their influence is given by the LDCA in $B^0$-decays, which with the aid of the corresponding measurement at the $B$-factories for $B^0_d$ only can be interpreted as an enhanced asymmetry in the $B_s^0$-system,
\begin{equation}
a_{sl}^s=\mbox{Im}\!\left(\frac{\Gamma_{12}^s}{M_{12}^s}\right)=\frac{|\Gamma_{12}^s|}{|M_{12}^s|}\,\sin{\phi_s}\, =\,
\frac{\Delta\Gamma_{B^0_s}}{\Delta M_{B^0_s}}\,\tan{\phi_s}\,,
\end{equation}
where $M_{12}^q -\frac{i}{2}\, \Gamma_{12}^q\equiv \langle B^0_q|\mathcal{H}^{\Delta B=2}_{\mathrm{eff}}|\bar{B}_q^0\rangle$ and $\phi_s={\rm arg}(-M_{12}^s/\Gamma_{12}^s)$.
Noting that in the \athdm the change in the rate difference is negligible, we can parametrize the relative influence of the charged scalar on the asymmetry as
\begin{equation}
\frac{\left.a_{sl}^s\right|_{\rm{full}}}{\left.a_{sl}^s\right|_{\rm{SM}}}=\frac{\sin\phi_s}{\sin\phi_s^{\rm SM}|\Delta_s|}\,,
\end{equation}
where $M_{12}^s|_{\rm full}=M_{12}^s|_{\rm SM}\Delta_s$. The possible range for this ratio is shown in Fig.~\ref{fig::LDCAnew} (details on the calculation can be found in \cite{Jung:2010ik}): the plot on the left hand side takes 
into account the constraint shown in Fig.~\ref{fig::udold}, varying all NP parameters in the ranges given before. However,
taking the correlations into account by using a parametrization $|\varsigma_u^*\varsigma_d|_{\rm max}(\varphi,M_{H^\pm})$ for the constraints shown in Fig.~\ref{fig::BRcorrelations}, the potential influence of charged-scalar effects on this observable is reduced by roughly a factor of 10 compared to the result without correlations. While an enhancement by a factor of $\sim5$ is still possible, larger effects from charged-scalar contributions are excluded.
This large impact can be understood as due to a combination of the two correlations shown above: as discussed in \cite{Jung:2010ik}, sizable values of the LDCA require relatively low charged-scalar masses, a sizable phase and large values for the product $|\varsigma_u^*\varsigma_d|$. The correlations in the constraint from $b\to s\gamma$ imply that the latter can only be achieved for relatively high masses and small values of the phase. Therefore the full suppression shows up only in the correlated analysis performed here.
A possible enhancement from additional neutral scalars remains untouched by this.
The present data suggest a preferred negative sign for the $B^0_s$ semileptonic asymmetry $a^s_{sl}$, which for charged-scalar effects to be in the right direction requires $\arg (\varsigma_u^*\varsigma_d^{\phantom{*}}) \in [3\pi/2,2\pi]$.
Due to the different structure of its amplitude, the $b\to s\gamma$ rate asymmetry could provide complementary information on the Yukawa phase. 

\section*{Note added}

After this paper was posted in the arXiv, a very relevant work on the rate asymmetry has appeared \cite{Benzke:2010tq}, showing that it receives a (parametrically leading in the SM) long-distance contribution arising from the interference of the electromagnetic dipole amplitude with an up-quark penguin transition accompanied by soft gluon emission.

\section*{Acknowledgements}
The authors would like to thank Mikolaj Misiak and Matthias Neubert for helpful discussions.
This work has been supported in part by the EU MRTN network FLAVIAnet [Contract No. MRTN-CT-2006-035482], by MICINN, Spain
[Grants FPA2007-60323 and Consolider-Ingenio 2010 Program CSD2007-00042 --CPAN--] and by Generalitat Valenciana [Prometeo/2008/069]. The work of P.T. is funded through an FPU Grant (MICINN, Spain).

\bibliography{bibliography}

\end{document}